\documentclass[aps,prd,twocolumn,superscriptaddress,nofootinbib]{revtex4-2}

\usepackage{amsmath,amssymb}
\usepackage{graphicx}
\usepackage{hyperref}
\usepackage{bm}

\begin{document}

\title{Catastrophe Structure and Stability of Axion Stars}

\author{Fedor V Kusmartsev}
\affiliation{Department of Physics, Khalifa University, Abu Dhabi, UAE}

\author{Joudy J F Beek}
\affiliation{Department of Physics, Khalifa University, Abu Dhabi, UAE}
\author{Jack C M Hughes}

\affiliation{Department of Public Health and Epidemiology, Khalifa University, Abu Dhabi, UAE}  
\affiliation{Department of Physics, Khalifa University, Abu Dhabi, UAE}

\date{\today}

\begin{abstract}
Axion stars,  which are self-gravitating Bose-Einstein condensates of QCD axions or axion-like particles, are compelling dark-matter candidates whose stability controls their cosmological and astrophysical impact. We develop a unified framework for axion-star stability that combines a non-relativistic Schr\"odinger--Poisson description with catastrophe theory. Starting from the Einstein--Klein--Gordon theory with a periodic axion potential, we derive the effective equations of motion and conserved quantities for axion stars and construct a one-parameter family of star-like solutions. Using a Gaussian variational ansatz, we obtain an analytic one-dimensional energy functional $E(R;N)$ and show that the appearance and disappearance of axion-star equilibria as a function of particle number $N$ is governed by Arnold $A_2$, $A_3$, and $A_4$ catastrophes. Including sextic self-interactions generates two distinct stable branches (dilute and dense axion stars)separated by an unstable branch. We further show how the catastrophe criteria extend beyond the sextic truncation to the full cosine and QCD-corrected axion potentials, for which the fold conditions are evaluated from the full variational energy. We distinguish explicitly between collective-coordinate stability within the ansatz and full radial perturbative stability, and we state the domain of validity of the non-relativistic approximation. Our results provide a compact classification of axion-star phases and identify model-dependent critical masses directly linked to collapse, bosenova-like emission, and black-hole formation.
\end{abstract}

\maketitle

\section{Introduction}

Axion stars sit at the intersection of particle physics, gravity and cosmology.
Axions were originally proposed to solve the strong–CP problem of QCD and have since emerged as leading cold–dark-matter candidates.
In many well–motivated cosmological histories, axions form a cold, highly occupied field that can condense gravitationally into self–bound Bose–Einstein condensates forming axion stars or, more generally, boson stars  bound by a balance of quantum pressure, self–gravity, and self-interactions~\cite{BraatenZhangColloquium,ZhangReview,Visinelli2018DiluteDense}. 
Cosmological simulations and semi-analytic arguments indicate\cite{buschmann2020early} that such objects form efficiently in the early Universe and in galactic halos, particularly during the ``dark ages,'' and may constitute a non-negligible fraction of dark matter.\cite{volovik2023gravity,volovik2003universe}

Axion stars are not just theoretical curiosities.
Their formation and evolution can influence the thermal and ionisation history of the Universe, alter small scale structure, and power potentially observable phenomena.
Collapsing axion stars may decay into photons, heating and ionising the intergalactic medium, with constraints emerging from CMB and 21\,cm data.\cite{PhysRevD.72.023501}
Encounters with neutron stars and magnetars have been proposed as engines for fast radio bursts, transient gamma ray emissions, and other multi-messenger signatures~\cite{Escudero2024,Acuna2025,Raby2016,Bai2016HAS}.
Compact axion stars or their remnants can also act as gravitational lenses or sources of gravitational waves(GW) in mergers,\cite{ramberg2019probing} offering multiple, complementary detection channels from microlensing surveys to radio arrays and GW interferometers~\cite{Escudero2024A,ChungJukko2024GW,Sugiyama2023,Chung-Jukko2023}.

At the heart of these possibilities lies a central open problem: \emph{the stability of axion stars across their parameter space}.
The equilibrium of an axion star is controlled by the competition between quantum pressure, self-gravity, and axion self-interactions.
For small central amplitudes, quantum pressure balances gravity, yielding ``dilute'' axion stars analogous to standard boson stars.\cite{KusmartsevMielkeSchunck1991BosonStarStability,KusmartsevMielkeSchunck1990NeutronBosonStability}. 
At larger amplitudes, attractive quartic interactions and repulsive higher order terms in the axion potential reshape the balance and allow for much more compact ``dense'' configurations~\cite{Visinelli2018DiluteDense,Braaten2016Dense}.
Numerical investigations have shown peaks in the mass–radius relationship and suggested the existence of several branches of stable configurations, separated by regions of instability, such as collapse. However, a simple, unified classification of these branches and of the transitions between them has remained elusive.

In this work we develop such a classification using \emph{catastrophe theory} as an organising principle.\cite{Arnold1992Catastrophe,Kusmartsev1989CatastropheMoleculesSolitons}
Starting from the relativistic Einstein-Klein-Gordon (EKG) theory with a periodic axion potential, we derive the non-relativistic Schr\"odinger-Poisson-Gross-Pitaevskii description appropriate for slowly moving axion condensates and identify the associated conserved quantities - particle number, total energy, momentum, and angular momentum - via Noether's theorem.
We then construct a one-dimensional variational energy functional $E(R;N)$ for axion stars using a Gaussian ansatz, in which the behaviour variable is the stellar radius $R$ and the primary control parameter is the axion number $N$.
This reduction allows us to apply Arnold's theory of singularities to the axion-star problem and to interpret the appearance, disappearance, and stability of equilibrium configurations as fold, cusp, and swallowtail catastrophes in the space of physical parameters\cite{Kusmartsev1989CatastropheMoleculesSolitons}.

The purpose of catastrophe theory here is not merely to rename extrema of a variational energy. It provides a global classification of the equilibrium manifold: it identifies where the map from internal solution parameters to conserved quantities loses rank, where stable and unstable branches are created or annihilated, and where critical particle numbers and maximum masses arise as folds or cusps of this map. In this sense, the catastrophe diagram complements the local variational minimum test by organizing the possible fates of the star across the full control-parameter space.

Finally, we confront this variational picture with a more faithful treatment of the axion potential.
By comparing polynomial truncations of the cosine (up to sextic and octic order) with the full periodic potential, we show explicitly when low order expansions are trustworthy and when they artificially introduce runaway collapse.
This clarifies the fate of dense axion stars and links analytic catastrophe theory arguments to the fully nonlinear dynamics observed in numerical simulations, including axion star ``bosenovae'' and black hole formation.

\section{Key Results}

\emph{Unified non-relativistic framework and conserved quantities.}  
Starting from the EKG system with a cosine axion potential, we show that in the non-relativistic, weak field regime, axion stars are described by the Schr\"odinger-Poisson-Gross-Pitaevskii system.
\begin{align}
 i \partial_t \psi &= -\frac{\nabla^2}{2m_a}\psi + m_a \Phi_N \psi + g |\psi|^2 \psi, \\
 \nabla^2 \Phi_N &= 4\pi G m_a |\psi|^2,
\end{align}
where $|\psi|^2$ is the axion number density, $\Phi_N$ the Newtonian potential, and $g$ the effective self--interaction coupling.
From the underlying Lagrangian, we derive, via Noether's theorem, the conserved axion number $N=\int |\psi|^2$, the total energy $E$, and, in general, the total momentum $P$ and angular momentum $L$.
For spherically symmetric, static stars, these reduce to $P=L=0$, leaving $(N,E)$ as the natural coordinates in which to classify equilibrium families and their stability.

\emph{Analytic fold catastrophe and critical mass for dilute axion stars.}  
Using a Gaussian density profile, we obtain a tractable one-dimensional energy functional
\begin{equation}
E(R;N) = \frac{kN}{R^2} - \frac{uN^2}{R} - \frac{sN^2}{R^3},
\label{eq:toy_quartic}
\end{equation}
with terms corresponding to quantum pressure ($+kN/R^2$), gravitational binding ($-uN^2/R$) and attractive quartic self-interaction ($-sN^2/R^3$).
Extremizing $E$ in $R$ yields two stationary radii—an outer minimum and an inner maximum—for axion numbers $N<N_c$, which coalesce at a critical value
\begin{equation}
N_c = \frac{k}{\sqrt{3us}},\qquad
R_c = \sqrt{\frac{3s}{u}}.
\end{equation}
Beyond $N_c$ no equilibrium exists.
The square root scaling $R_\pm(N)\simeq R_c\pm C\sqrt{N_c-N}$ near this point identifies it as a fold (saddle node) catastrophe and provides a simple analytic expression for the maximum mass of dilute axion stars and the onset of collapse in terms of microscopic parameters.

\emph{Sextic self-interaction and the emergence of dilute and dense branches.}  
Including the next-order, repulsive sextic interaction leads to the extended energy functional
\begin{equation}
E(R;N) = \frac{kN}{R^2} - \frac{uN^2}{R} - \frac{sN^2}{R^3} + \frac{dN^3}{R^6},
\label{eq:toy_sextic}
\end{equation}
where $d>0$ parameterises effective three-body repulsion.
We show that this modification yields a double-branch structure:
for small $N$ there is a single large $R$ minimum (dilute star);
for intermediate $N$ the energy develops two minima (dilute and dense stars) and one maximum (unstable branch);
for larger $N$ the dilute minimum disappears in a second fold, leaving only the dense branch.
Solving the degeneracy conditions $\partial_R E=\partial_R^2 E=0$, we derive a cusp surface in the $(N,\beta)$ plane, where $\beta=192dku/s^3$, and identify the corresponding phase diagram with regions of dilute stability, dense stability, and no equilibrium.

\emph{Swallowtail catastrophe and multi-branch stability.}  
When both $N$ and the effective self-interaction strength are treated as control parameters, the two cusps just described join into an Arnold $A_4$ swallowtail catastrophe.
Slices of the catastrophe surface at fixed couplings exhibit up to three physical equilibria:
a small $R$ stable branch (dense stars), an intermediate unstable branch, and a large $R$ stable branch (dilute stars).
The two folds and corresponding cusps of the swallowtail indicate where the dilute and dense branches vanish, respectively, and they correspond to critical masses at which slowly evolving axion stars are forced to undergo abrupt transitions.
\emph{Full cosine potential removes artificial runaways and renders dense stars metastable.}  
Replacing the polynomial approximation with the full periodic potential $V(\phi)=m_a^2 f_a^2[1-\cos(\phi/f_a)]$ in the Gaussian model, we show that the total energy always diverges upward, $E(R;N)\to +\infty$, as $R\to 0$, dominated by the gradient term.
In contrast, an 8th-order truncation of the cosine produces $E(R;N)\to -\infty$ at small $R$ because of the negative $\phi^8$ term, leading to an unphysical runaway collapse.
The faithful cosine treatment therefore regularizes the inner behavior: dense stars appear as metastable minima of a globally bounded energy landscape and, when pushed past the folds and their cusps of the swallowtail, evolve dynamically into bosenova-like explosions, black holes, or dispersal, in agreement with numerical EKG simulations.

\emph{Catastrophe theory phase diagram and observational stakes.}  
Together these results define a catastrophe-theory phase diagram for axion stars.
The locations of the folds and cusps in $(N,f_a,m_a,g,\dots)$ space specify the maximum mass of stable dilute stars, the parameter range where dense stars are long-lived, and the thresholds beyond which no static equilibria exist.
This diagram translates directly into constraints on axion models if axion stars are to survive in galaxies today, and into predictions for the rates and energies of axion star explosions, compact mergers, and associated gravitational wave and radio signals.

\section{Theoretical Framework}
\label{sec:TheoreticalFramework}

\subsection{Relativistic formulation, axion potential, and non-relativistic limit}

Our starting point is the relativistic Einstein-Klein-Gordon (EKG) system for a real axion field
$\phi$ with a periodic self–interaction.
Following Mielke's soliton model of dark matter and natural inflation~\cite{Mielke2019SolitonDMNatInflation},
we view the axion as a scalar field obeying a nonlinear Klein-Gordon (NLKG) equation
\begin{equation}
\square \phi = \frac{\partial U(\phi)}{\partial \phi},
\label{eq:NLKG_general}
\end{equation}
where $\square$ is the covariant d'Alembertian and $U(\phi)$ is an effective potential.
For QCD axions or axion-like particles the potential is periodic,
\begin{equation}
U_a(\phi) = m_a^2 f_a^2\left[1-\cos\!\left(\frac{\phi}{f_a}\right)\right],
\label{eq:axion_potential_cos}
\end{equation}
with axion mass $m_a$ and decay constant $f_a$.
In the small-field regime $|\phi|\ll f_a$, Eq.~\eqref{eq:axion_potential_cos} reduces to
the familiar quadratic plus attractive quartic and repulsive sextic interactions,
whereas at large amplitude the full cosine becomes essential~\cite{Mielke2019SolitonDMNatInflation}.

In flat $(1+1)$ dimensional spacetime, the NLKG equation with the periodic potential
\eqref{eq:axion_potential_cos} is precisely the sine--Gordon system,
whose kink and breather solutions provide a prototypical example of relativistic solitons.
Mielke has shown that kink-kink and kink-antikink collisions in this model can be
constructed analytically using B\"acklund transformations, and that at late times the
solitons re–emerge with their shapes and velocities essentially intact, apart from small
inelastic effects at relativistic energies~\cite{Mielke2019SolitonDMNatInflation}.
These $(1+1)$D solitons serve as a useful toy model for the ``lumps'' that appear in higher
dimensions, where gravity must participate in stabilisation.

In $(3+1)$ dimensions with gravity, the axion field is minimally coupled to Einstein gravity via the action
\begin{equation}
S = \int d^4x\,\sqrt{-g}\,\mathcal{L}, \qquad
\mathcal{L} = \frac{M_{\rm Pl}^2}{2}R
-\frac{1}{2}\partial_\mu\phi\,\partial^\mu\phi
-U_a(\phi),
\label{eq:action_EKG}
\end{equation}
where $M_{\rm Pl}$ is the reduced Planck mass and $R$ the Ricci scalar.
Varying with respect to the metric and the scalar field yields Einstein's equations with
the axion energy-momentum tensor and the curved space NLKG equation
\begin{equation}
\square\phi + m_a^2 f_a \sin\!\left(\frac{\phi}{f_a}\right)=0.
\label{eq:EKG_cosine}
\end{equation}

More general spherically symmetric, self-gravitating solutions of this system such as oscillatons or axion stars may be constructed numerically. Indeed such approaches have been used in a number of works, particularly in the dilute regime where
$|\phi|\ll f_a$ and $U_a(\phi)$ is approximately quadratic.\cite{BraatenZhangColloquium}

An alternative, complementary viewpoint advocated by Mielke and co-workers is to
Legendre transform the scalar sector into an effective higher - order curvature theory
with Lagrangian $L(R)$, where $R$ is the scalar curvature~\cite{Benitez1997COBEtoCurvature,
MielkeKusmartsevSchunck2008InflationCurvature,Mielke2019SolitonDMNatInflation}.
In this geometric picture the scalar field plays the role of a control parameter and the
Whitney surface $L(R)$ acquires a cusp or swallowtail structure for realistic axion type
potentials.  For the cosine potential~\eqref{eq:axion_potential_cos} used in natural inflation,
Mielke finds that $L(R)$ exhibits a pronounced swallowtail catastrophe whose branches
correspond to distinct cosmological phases, with the energetically lowest branch inducing
an effective cosmological constant~\cite{Mielke2019SolitonDMNatInflation}.
Our catastrophe theory analysis of axion stars can be viewed as a ``microscopic''
counterpart of this curvature-based construction: instead of $L(R)$ we work with an effective
energy $E(R;N)$ in which the stellar radius $R$ is the behaviour variable and the
axion number $N$ and couplings play the role of control parameters.

For the parameter range relevant to galactic and cosmological axion stars, the bulk of each
configuration is non-relativistic and only weakly self-gravitating.
It is therefore convenient to introduce a complex field $\psi$ via the standard expansion
\begin{equation}
\phi(\bm{x},t) \simeq \sqrt{\frac{1}{2m_a}}
\left(\psi\,e^{-im_a t} + \psi^\ast e^{+im_a t}\right),
\label{eq:NR_decomposition}
\end{equation}
and to average over the fast oscillations at frequency $m_a$.
To leading order in $v/c$ and in the weak field limit, this procedure yields the
non-relativistic Schr\"odinger-Poisson-Gross-Pitaevskii system of equations, \cite{cazenave1998introduction}
which we take as the starting point for our
variational and catastrophe theory analysis of axion stars.


\subsection{Stationary, spherically symmetric axion stars}

Star-like configurations correspond to time-harmonic, spherically symmetric solutions
\begin{equation}
\psi(\bm{x},t)=e^{-i\mu t}\psi_0(r),
\end{equation}
where $\mu$ plays the role of a chemical potential and $\psi_0(r)$ is real and positive.\cite{KusmartsevMielkeSchunck1991BosonStarStability,KusmartsevMielkeSchunck1990NeutronBosonStability}
Inserting this ansatz into the Schr\"odinger-Poisson equations yields the coupled radial equations
\begin{equation}
-\frac{1}{2m_a}\frac{1}{r^2}\frac{d}{dr}\!\left(r^2\frac{d\psi_0}{dr}\right)
+ m_a \Phi_N(r)\psi_0(r) + g\psi_0^3(r)=\mu\psi_0(r)
\end{equation}
\begin{equation}
\frac{1}{r^2}\frac{d}{dr}\!\left(r^2\frac{d\Phi_N}{dr}\right) = 4\pi G m_a \psi_0^2(r),
\end{equation}
with regularity conditions $d\psi_0/dr|_{r=0}=d\Phi_N/dr|_{r=0}=0$ and localisation $\psi_0,\Phi_N\to 0$ as $r\to\infty$.
This one-parameter family of solutions, which depends also on the value of central density is characterised by eigenvalue $\mu$.  Each solution defines the density-radius curve of axion stars in the non-relativistic regime.
The star  stability we will  determine with the aid of the  catastrophe theory.\cite{Kusmartsev1989CatastropheMoleculesSolitons}

\subsection{Lagrangian formulation and conserved quantities}

The non-relativistic equations follow from the Lagrangian density
\begin{align}
\mathcal{L} &= \frac{i}{2}\bigl( \psi^\ast \partial_t\psi - \psi\,\partial_t\psi^\ast \bigr)
-\frac{|\nabla\psi|^2}{2m_a} - m_a \Phi_N |\psi|^2\nonumber\\
&\quad - \frac{g}{2}|\psi|^4 + \frac{1}{8\pi G}|\nabla\Phi_N|^2.
\end{align}
A global phase symmetry $\psi\to e^{i\alpha}\psi$ yields conservation of the axion number
\begin{equation}
N = \int d^3x\,|\psi|^2 = 4\pi \int_0^\infty dr\,r^2 \psi_0^2(r),
\end{equation}
while time--translation invariance yields conservation of the total energy
\begin{equation}
E=\int d^3x\left[\frac{|\nabla\psi|^2}{2m_a} + m_a\Phi_N|\psi|^2 + \frac{g}{2}|\psi|^4 + \frac{|\nabla\Phi_N|^2}{8\pi G}\right].
\end{equation}
Spatial translations and rotations imply conservation of momentum and angular momentum; these vanish for non-rotating, spherically symmetric stars, leaving $(N,E)$ as the key macroscopic invariants.
Stability changes occur when the mapping from internal parameters (e.g.\ $\mu$ or central field value) to $(N,E)$ becomes singular, precisely the setting where catastrophe theory applies.\cite{Kusmartsev1989CatastropheMoleculesSolitons,KusmartsevMielkeSchunck1991BosonStarStability,KusmartsevMielkeSchunck1990NeutronBosonStability}

\section{Catastrophe Theory for Molecules and Stars}
\label{sec:catastrophe-molecules-stars}

The mathematical language underlying our analysis is Arnold's catastrophe theory, which classifies the qualitative changes of equilibria in terms of a small set of elementary singularities such as folds, cusps and swallowtails~\cite{Arnold1992Catastrophe}.
The basic ingredients are:
(i) a behaviour variable $x$ (or a small number of such variables), which parameterises the possible configurations of the system;
(ii) a set of control parameters $\bm{a}$ (external fields, couplings, particle number, \emph{etc.});
and (iii) an effective potential $V(x;\bm{a})$ whose stationary points describe equilibria.
Catastrophes occur where the map from internal parameters to observables loses invertibility, i.e.\ where stationary points become degenerate and are created or annihilated.

Kusmartsev and collaborators were the first to apply this framework systematically to nonlinear field theories and many-body systems.
In molecular physics they showed that the reorganisation of electronic states and chemical bonds can be described by fold, cusp and higher-order catastrophes of an effective potential defined on a low--dimensional nuclear configuration space~\cite{Kusmartsev1989CatastropheMoleculesSolitons}.
As external parameters such as inter-atomic distances or electric fields are varied, the molecular energy surface develops characteristic cusp and swallowtail structures, which organise phenomena such as isomerisation, level crossings and sudden changes of stability.
Related constructions for soliton-bearing field theories (sine-Gordon, $\phi^4$, $\phi^6$, \emph{etc.}) exploit exact and approximate solutions obtained by B\"acklund and Darboux transformations~\cite{RogersSchief2002BacklundDarboux,Gulevich2009JosephsonShapeWobbling,Khare2014Phi8Phi10Phi12}, revealing how families of kinks, breathers and their bound states are connected through catastrophe surfaces in the relevant control spaces~\cite{Kusmartsev1989CatastropheMoleculesSolitons}. The method can also be applied to more complicated structures, such as orbital glass consisting of arbitrarily oriented vortices originated from the flux cloning \cite{kusmartsev1992orbital,Gulevich2006,Sutton2013}.

The same methodology has been extended to self-gravitating systems.
In a series of papers, Kusmartsev, Mielke and Schunck applied catastrophe theory to neutron stars and boson stars~\cite{KusmartsevMielkeSchunck1990NeutronBosonStability,KusmartsevMielkeSchunck1991BosonStarStability,SchunckKusmartsevMielke1992NumericalRelativity}.
There the behaviour variables are global properties of the configuration, such as the stellar radius or central density, while the control parameters include the total particle number and coupling constants in the matter sector.
Mass-radius and mass-particle-number diagrams for boson stars exhibit folds and cusps, delimiting regions of stability, metastability and inevitable collapse.
Crossing a fold corresponds to the gain or loss of a single unstable mode, while cusp points mark the junction of different stability branches.
This programme demonstrated that boson stars possess an extended stable branch with astrophysically interesting masses and radii, as well as more exotic rotating and toroidal configurations whose stability can also be classified catastrophically~\cite{MielkeSchunck2000BosonStarsPBH,Mielke2016RotatingBosonStars}.

 Mielke and co-workers later pushed this idea further by mapping scalar-field models to purely gravitational theories with higher-order curvature Lagrangians $L(R)$ via a Legendre transform and conformal rescaling~\cite{Benitez1997COBEtoCurvature,SchunckKusmartsevMielke2005DMandCurvature,MielkeKusmartsevSchunck2008InflationCurvature,MielkeRomero2006Quintaxion}.
In this representation the scalar field plays the role of a control parameter, while the behaviour variable is the scalar curvature $R$.
The resulting function $L(R)$ defines a so-called Whitney surface whose valleys and ridges encode the bifurcations of the underlying scalar-gravity system.
For simple toy models, one obtains semi-cubic cusps in the $(R,L)$ plane, while for realistic inflationary potentials, the surface can develop swallowtail structures.
In particular, Mielke has shown that an axion-type cosine potential used in natural inflation generates a higher-order curvature Lagrangian $L(R)$ with a pronounced swallowtail catastrophe~\cite{Mielke2019SolitonDMNatInflation}: different branches of $L(R)$ correspond to distinct cosmological phases, and the lowest branch picks out an effective cosmological constant, linking axion-like fields, dark matter and dark energy within a single geometric picture.

Our treatment of axion stars fits naturally into this catastrophe-theory lineage.
The Gaussian energy functional $E(R;N)$ introduced in Sec.~\ref{sec:VariationalModel} plays the role of the effective potential; the stellar radius $R$ is the behavior variable, and the control parameters are the axion number $N$ and the self-interaction strengths inherited from the microscopic axion potential.
The fold, cusp, and swallowtail structures we identify in $(N,E)$ space for axion stars are thus the direct analogs of the catastrophes previously found in molecular systems and boson stars by Kusmartsev, Mielke, and collaborators~\cite{Kusmartsev1989CatastropheMoleculesSolitons,KusmartsevMielkeSchunck1991BosonStarStability,MielkeKusmartsevSchunck2008InflationCurvature,Mielke2019SolitonDMNatInflation}.
In the remainder of the paper, we make this correspondence explicit and show how the axion-star phase diagram can be read as a concrete realization of these universal catastrophe surfaces.


\section{Variational Model and Catastrophe Structure}
\label{sec:VariationalModel}

\subsection{Gaussian ansatz and explicit coefficients}
\label{sec:variational_coefficients}

To gain analytic control we adopt the normalized spherical Gaussian ansatz
\begin{equation}
\psi(r)=\left(\frac{N}{\pi^{3/2}R^3}\right)^{1/2}
\exp\left(-\frac{r^2}{2R^2}\right),
\label{eq:gaussian_ansatz_revised}
\end{equation}
where $N=\int d^3x\,|\psi|^2$ and $R$ is the collective radius.  In the non-relativistic Gross--Pitaevskii limit we write the self-interaction energy as
\begin{equation}
E_{\rm SI}=\frac{g_2}{2}\int d^3x\,|\psi|^4
+\frac{g_3}{3}\int d^3x\,|\psi|^6+\cdots,
\label{eq:SI_convention_revised}
\end{equation}
where $g_2<0$ represents the leading attractive two-body interaction and $g_3>0$ the leading repulsive three-body interaction generated by higher-order terms in the axion potential.  For the Gaussian profile the variational energy takes the form
\begin{equation}
E(R;N)=\frac{kN}{R^2}-\frac{uN^2}{R}-\frac{sN^2}{R^3}+\frac{dN^3}{R^6},
\label{eq:energy_kusd_revised}
\end{equation}
with
\begin{equation}
k=\frac{3}{4m_a},
\qquad
u=\frac{Gm_a^2}{\sqrt{2\pi}},
\label{eq:k_u_revised}
\end{equation}
\begin{equation}
s=\frac{|g_2|}{4\sqrt{2}\,\pi^{3/2}},
\qquad
d=\frac{g_3}{3^{5/2}\pi^3}.
\label{eq:s_d_revised}
\end{equation}
Here the four terms in Eq.~\eqref{eq:energy_kusd_revised} are, respectively, quantum pressure, Newtonian gravitational binding, attractive quartic self-interaction, and repulsive sextic self-interaction.
For the cosine potential
\begin{equation}
V(\phi)=m_a^2f_a^2\left[1-\cos\left(\frac{\phi}{f_a}\right)\right],
\label{eq:cosine_potential_revised}
\end{equation}
the leading non-relativistic coefficients are
\begin{equation}
g_2\simeq -\frac{1}{8f_a^2},
\qquad
g_3\simeq \frac{1}{96m_af_a^4},
\label{eq:g2_g3_revised}
\end{equation}
so that
\begin{equation}
s\simeq \frac{1}{32\sqrt{2}\,\pi^{3/2}f_a^2},
\qquad
d\simeq\frac{1}{96\,3^{5/2}\pi^3m_af_a^4}.
\label{eq:s_d_axion_revised}
\end{equation}
These relations make explicit the connection between the macroscopic variational parameters and the microscopic axion parameters $(m_a,f_a)$, together with Newton's constant $G$.

\subsection{Dependence on the variational ansatz}
\label{sec:ansatz_dependence}

The Gaussian ansatz is used because it gives closed analytic expressions for all energy contributions and therefore exposes the catastrophe structure transparently.  The catastrophe classification is not tied to this particular profile.  For a general normalized one-parameter profile
\begin{equation}
\psi_\chi(r;R)=N^{1/2}R^{-3/2}\chi(r/R),
\end{equation}
where $ \chi(x)$ is normalized wave function:
\begin{equation}
4\pi\int_0^\infty dx\,x^2|\chi(x)|^2=1,
\label{eq:general_profile_revised}
\end{equation}
then, the same dimensional scalings give
\begin{equation}
E_\chi(R;N)=\frac{k_\chi N}{R^2}-\frac{u_\chi N^2}{R}
-\frac{s_\chi N^2}{R^3}+\frac{d_\chi N^3}{R^6}+\cdots,
\label{eq:general_ansatz_energy_revised}
\end{equation}
where the numerical shape factors $k_\chi,u_\chi,s_\chi,d_\chi$ depend on the chosen function $\chi$.  Thus Gaussian, exponential, or exponential-linear trial functions shift the numerical values of the critical masses and radii, but they do not change the fold/cusp topology as long as the same signs and powers of $R$ appear in the effective energy. This includes the exponential-linear variational profiles often used in axion-star microlensing studies, such as those considered by Fujikura \emph{et al.}~\cite{Fujikura2021Microlensing}; such choices renormalize the shape factors but leave the catastrophe topology unchanged.

\subsection{Quartic model and the analytic fold}

The leading quartic model is obtained by dropping the sextic term in Eq.~\eqref{eq:energy_kusd_revised}:
\begin{equation}
E(R;N)=\frac{kN}{R^2}-\frac{uN^2}{R}-\frac{sN^2}{R^3}.
\label{eq:toy_quartic_revised}
\end{equation}
Extrema satisfy
\begin{equation}
\frac{\partial E}{\partial R}= -\frac{2kN}{R^3}+\frac{uN^2}{R^2}+\frac{3sN^2}{R^4}=0,
\end{equation}
or equivalently
\begin{equation}
uN R^2-2kR+3sN=0.
\end{equation}
This quadratic has two positive roots for $N<N_c$, one double root at $N=N_c$, and no real positive root above it, with
\begin{equation}
N_c=\frac{k}{\sqrt{3us}},
\qquad
R_c=\sqrt{\frac{3s}{u}}.
\end{equation}
The coalescence at $(R_c,N_c)$ is the canonical fold ($A_2$) catastrophe and gives the variational maximum mass of the dilute branch.

\subsection{Sextic self-interaction, exact control parameter, and cusp surface}

Including the repulsive sextic self-interaction gives Eq.~\eqref{eq:energy_kusd_revised}.  Equilibria satisfy
\begin{equation}
\frac{\partial E}{\partial R}= -\frac{2kN}{R^3}+\frac{uN^2}{R^2}+\frac{3sN^2}{R^4}-\frac{6dN^3}{R^7}=0,
\end{equation}
which may be written as
\begin{equation}
uN R^5-2kR^4+3sNR^3-6dN^2=0.
\label{eq:quintic_revised}
\end{equation}
The fold set is defined by
\begin{equation}
\partial_R E=0,
\qquad
\partial_R^2E=0,
\label{eq:fold_conditions_revised}
\end{equation}
and a cusp in an enlarged control space additionally satisfies $\partial_R^3E=0$. Solving the fold conditions and introducing
\begin{equation}
x=\frac{uR^2}{s},
\qquad
\boxed{\beta=\frac{192\,dku}{s^3}},
\label{eq:beta_exact_revised}
\end{equation}
yields the dimensionless cubic
\begin{equation}
5x^3-6x^2-27x+\beta=0.
\label{eq:cubic_beta_revised}
\end{equation}
The two positive roots of Eq.~\eqref{eq:cubic_beta_revised}, when present, correspond to the two folds that bound the region where three extrema coexist: a dilute minimum, an intermediate maximum, and a dense minimum.  This is the $A_3$ cusp cross-section of the equilibrium manifold.

\subsection{Swallowtail catastrophe and hysteresis}

\begin{figure}[t]
\centering
\includegraphics[width=0.95\linewidth]{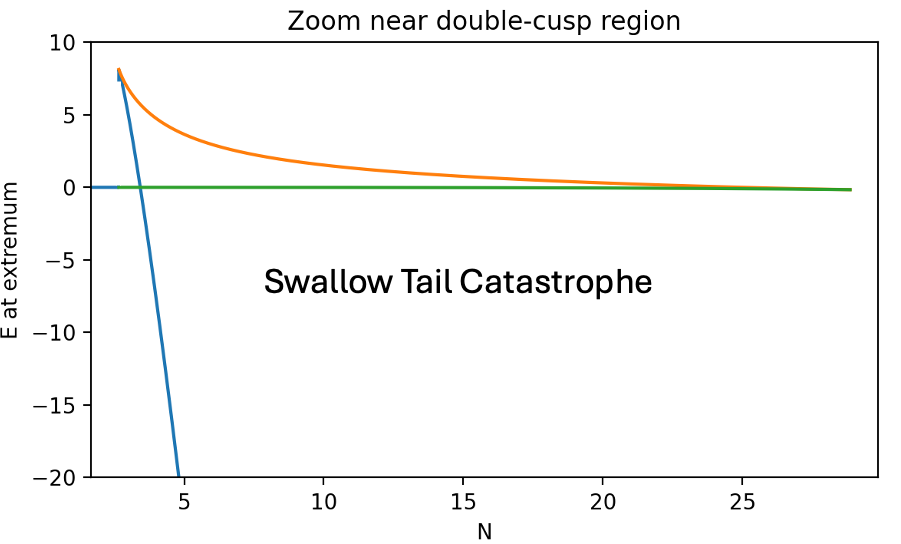}
\caption{Energy at extrema as a function of axion number $N$ for a representative sextic toy model. Blue: small-$R$ stable branch (dense stars). Red: unstable branch. Green: large-$R$ stable branch (dilute stars). The two cusps delimit the region where all three branches coexist; in the full control space this region is part of an Arnold $A_4$ swallowtail catastrophe.}
\label{fig:swallowtail}
\end{figure}

When both $N$ and the effective self-interaction strength are varied, the two cusps connect into an Arnold $A_4$ swallowtail catastrophe.  Figure~\ref{fig:swallowtail} shows a representative slice of this surface. Three branches of extrema are visible: a small-$R$ stable branch (dense stars), a middle-$R$ unstable branch, and a large-$R$ stable branch (dilute stars).  Crossing a fold forces a rapid reconfiguration and can drive collapse, bosenova-like emission, dispersal, or black-hole formation depending on the relativistic dynamics beyond the variational regime.

\begin{figure}[t]
\centering
\includegraphics[width=1.0\linewidth]{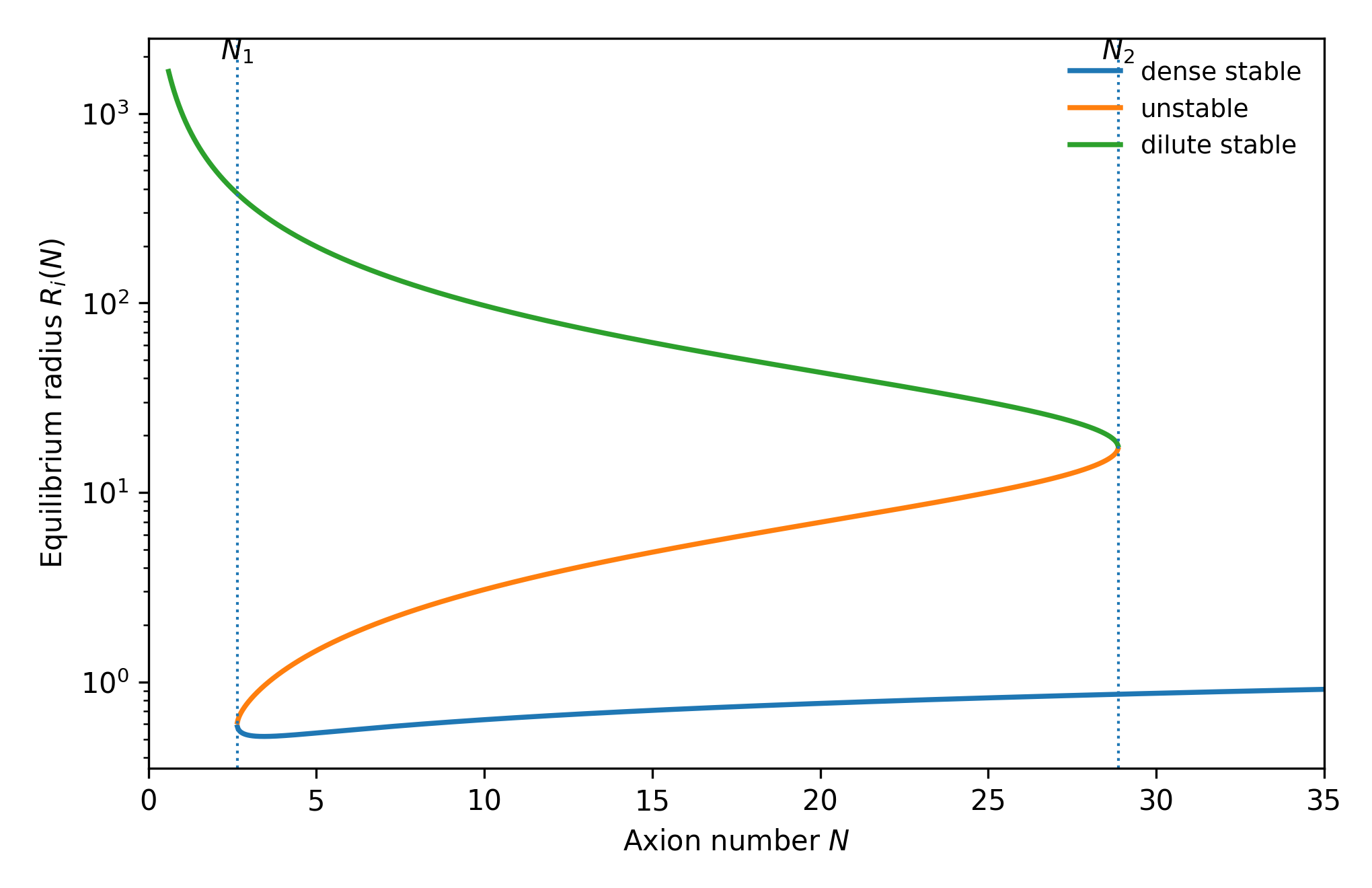}
\caption{%
\textbf{Equilibrium branches and cusp structure in the sextic variational model.}
Equilibrium radii of the Gaussian axion-star ansatz as a function of the axion number $N$, obtained by solving
$\partial_R E(R;N)=0$ for the sextic effective energy
$E(R;N)=kN/R^2-uN^2/R-sN^2/R^3+dN^3/R^6$.
The plot uses the representative dimensionless parameters $k=5$, $u=0.01$, $s=1$, and $d=0.01$, for which
$\beta=192dku/s^3=0.096$.
For each value of $N$, all real positive solutions of
$uNR^5-2kR^4+3sNR^3-6dN^2=0$ are retained and classified by the collective-coordinate stability criterion $\partial_R^2E>0$ (stable variational minimum) or $\partial_R^2E<0$ (unstable maximum).
The large-$R$ stable branch corresponds to dilute axion stars, the small-$R$ stable branch to dense axion-star configurations, and the intermediate branch to an unstable saddle separating them.
The two turning points, marked by the fold conditions $\partial_R E=\partial_R^2E=0$, occur at approximately $(N_1,R_1)\simeq(2.64,0.60)$ and $(N_2,R_2)\simeq(28.87,17.32)$.
Between these folds the variational energy has three extrema, giving the coexistence region of dilute and dense stable configurations separated by an unstable branch. Outside this interval only one equilibrium branch remains.
This is the one-dimensional cross-section of the Arnold $A_3$ cusp structure; allowing an additional control parameter unfolds it into the swallowtail-type phase structure discussed in the text.
}
\label{fig:branch_diagram}
\end{figure}


\section{Full Axion Potential and Breakdown of Polynomial Truncations}
\label{sec:FullPotential}

The catastrophe structure derived above is not restricted to a polynomial truncation.  For any variational energy $E(R;N)$ the equilibrium manifold and its folds are defined by
\begin{equation}
\partial_R E(R;N)=0,
\qquad
\partial_R^2E(R;N)=0.
\label{eq:full_fold_general}
\end{equation}
The sextic model is useful because these equations can be solved analytically, but the same criterion applies to the full axion potential once the potential-energy integral is evaluated numerically.

For the exact cosine potential the variational energy is
\begin{equation}
E_{\rm full}(R;N)=E_{\rm grad}+E_{\rm grav}
+V(\phi).
\label{eq:E_full_revised}
\end{equation}

where
\begin{equation}
V(\phi)=\int d^3x\,m_a^2f_a^2\left[1-\cos\left(\frac{\phi(r;R,N)}{f_a}\right)\right].
\end{equation}
Folds and cusps are obtained from Eq.~\eqref{eq:full_fold_general}, generally by numerical evaluation.  Thus the full potential changes the numerical location of the critical points, but it does not change the catastrophe-theory construction.

In the small-field regime the cosine expands as
\begin{equation}
V(\phi)\simeq
\frac12 m_a^2\phi^2
-\frac{m_a^2}{4!f_a^2}\phi^4
+\frac{m_a^2}{6!f_a^4}\phi^6
-\frac{m_a^2}{8!f_a^6}\phi^8+\cdots.
\label{eq:taylor_revised}
\end{equation}
For the Gaussian ansatz, increasingly high powers of $\phi$ generate increasingly high inverse powers of $R$ as $R$ decreases.  The sextic truncation has a positive highest-power contribution and therefore produces an artificial hard wall at small $R$, whereas the octic truncation ends on a negative term and produces an unphysical runaway $E_8\to-\infty$.  The full cosine potential is bounded, $0\le V\le 2m_a^2f_a^2$, and therefore avoids this truncation artifact.  However, the limit $R\to0$ lies outside the non-relativistic domain; in this paper it is used only as a diagnostic of whether a truncation has introduced an unphysical variational direction.

\begin{figure}[t]
\centering
\includegraphics[width=1.00\linewidth]{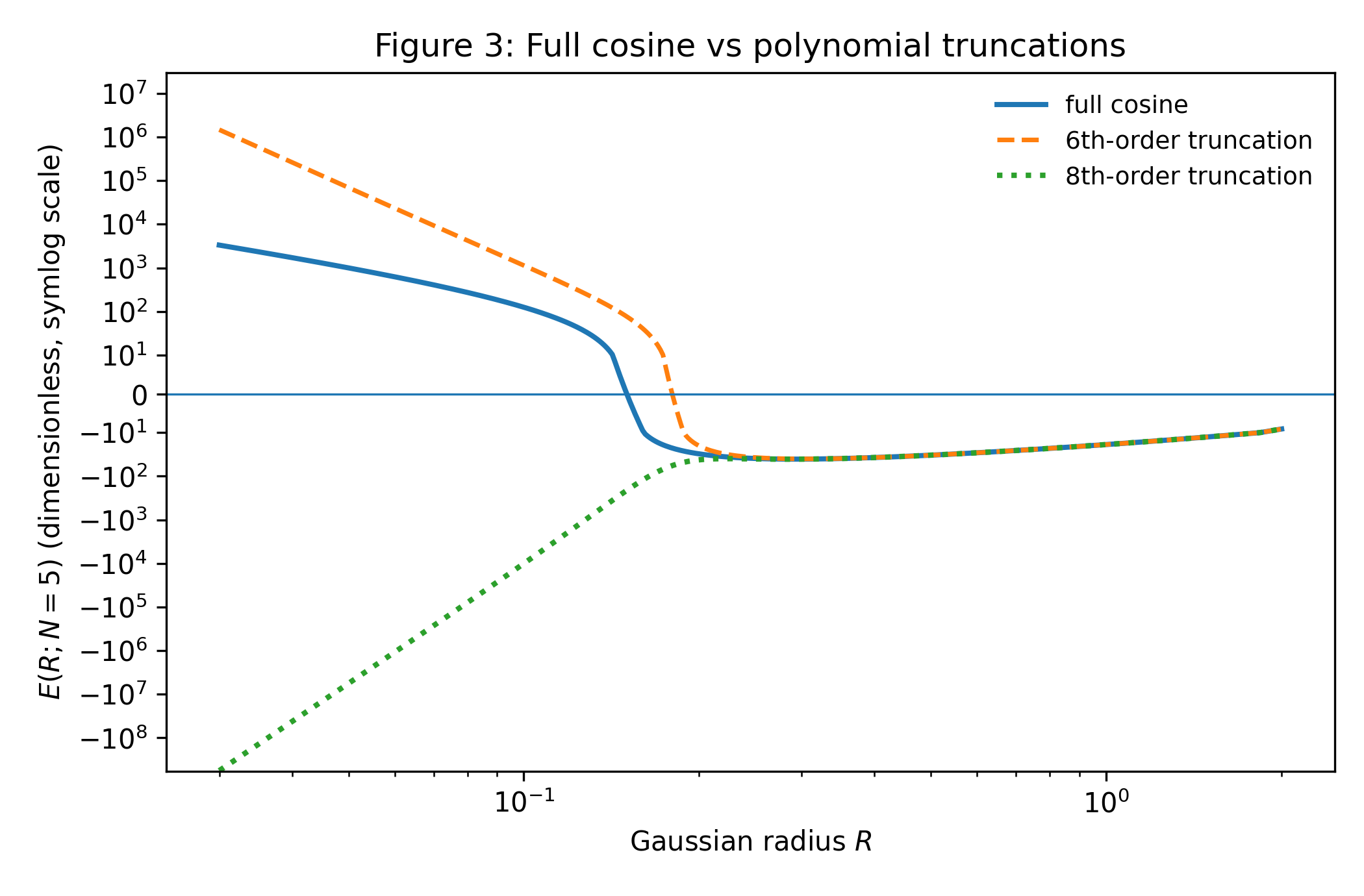}
\caption{%
\textbf{Full cosine potential versus polynomial truncations in the Gaussian variational energy.}
Comparison of the effective energy $E(R;N)$ for a fixed dimensionless axion number $N=5$, using the same Gaussian real-field profile
$\phi(r)=\phi_0\exp[-r^2/(2R^2)]$, with $\phi_0^2=N/(\pi^{3/2}R^3)$.
The total energy contains gradient, gravitational, and potential terms, $E(R;N)=E_{\rm grad}+E_{\rm grav}+E_{\rm pot}$, with the representative dimensionless choices $k_{\rm grad}=0.75$, $u_{\rm grav}=1$, and $m=f=1$.
The potential contribution is evaluated numerically for three cases: $V_{\rm full}=1-\cos\phi$, $V_6=\phi^2/2-\phi^4/4!+\phi^6/6!$, and $V_8=V_6-\phi^8/8!$.
The full cosine potential remains bounded and gives an energy profile that rises at sufficiently small $R$ within the variational model, while the sextic truncation mimics this behaviour by ending on a positive $\phi^6$ term. By contrast, the octic truncation ends on the negative $-\phi^8$ contribution and produces an artificial runaway direction, $E\to-\infty$, at small $R$.
This illustrates why finite polynomial truncations can over-stabilize or over-destabilize dense axion-star configurations, depending on the sign of the highest retained term. The small-$R$ region is shown only as a diagnostic of the variational potential and of truncation artifacts; the non-relativistic description itself is not valid arbitrarily close to $R=0$.
}
\label{fig:full_vs_trunc}
\end{figure}

\subsection{QCD-corrected axion potential}
\label{sec:qcd_corrected_potential}

The cosine potential is an analytically transparent reference model.  For the QCD axion, chiral perturbation theory gives a more precise potential, which may be written as
\begin{equation}
V_{\rm QCD}(\theta)=m_\pi^2f_\pi^2
\left[
1-
\sqrt{1-\frac{4z}{(1+z)^2}\sin^2\left(\frac{\theta}{2}\right)}
\right],
\label{eq:qcd_potential_revised}
\end{equation}
where $\theta=\frac{\phi}{f_a}$, and $z=m_u/m_d$.  Using
\begin{equation}
m_a^2f_a^2=m_\pi^2f_\pi^2\frac{z}{(1+z)^2},
\end{equation}
this can equivalently be written as
\begin{equation}
V_{\rm QCD}(\theta)=m_a^2f_a^2\frac{(1+z)^2}{z}
\left[
1-
\sqrt{1-\frac{4z}{(1+z)^2}\sin^2\left(\frac{\theta}{2}\right)}
\right].
\label{eq:qcd_potential_mafa_revised}
\end{equation}
This potential modifies the numerical quartic, sextic, and higher-order coefficients.  The catastrophe criterion remains the same: one computes the full variational energy with $V_{\rm QCD}$ and solves Eq.~\eqref{eq:full_fold_general}.  Thus the fold/cusp classification is a property of the equilibrium manifold and not of a particular Taylor truncation. See Ref.~\cite{GrilliDiCortona2016} for the precise QCD axion potential.


\section{Variational Stability and Domain of Validity}
\label{sec:stability_validity}

\subsection{Collective-coordinate stability versus radial perturbation stability}
\label{sec:variational_vs_radial}

The condition $\partial_R^2E(R;N)>0$ establishes stability only with respect to the collective coordinate $R$ inside the chosen variational family.  It means that the trial configuration is a local minimum against changes of its overall size within the ansatz.  Full radial stability is a stronger statement: it requires the second variation of the energy functional at fixed $N$, or equivalently the spectrum of radial perturbations about the equilibrium solution.  Catastrophe theory locates turning points where the number of unstable modes may change, but it does not replace the full radial eigenvalue problem.  Accordingly, throughout this work ``stable'' refers to variational or branch stability unless a full perturbative statement is explicitly specified.

\subsection{Non-relativistic validity and the small-radius limit}
\label{sec:nr_validity}

The non-relativistic variational description assumes
\begin{equation}
R\gg m_a^{-1},
\qquad
|\Phi_N|\ll1,
\qquad
v\ll 1,
\end{equation}
and a binding energy per axion much smaller than $m_a$.  The formal $R\to0$ behaviour of a polynomial variational energy therefore should not be interpreted as a physical prediction of the final collapse endpoint.  We use this limit only to diagnose the mathematical behaviour of approximate potentials: a finite truncation ending on a negative highest-order term may create an artificial unbounded direction that is absent in the full periodic potential.  The final fate of a dense configuration beyond the non-relativistic regime must be determined from the relativistic Einstein--Klein--Gordon evolution.


\section{Astrophysical Implications}

The catastrophe theory phase diagram for axion stars provides a direct bridge between microscopic axion physics and macroscopic, potentially observable phenomena.

The dimensionless parameters $k,u,s,d$ in our variational model can be expressed in terms of $m_a$, $f_a$ and $G$ using explicit Gaussian integrals.
The critical axion numbers $N_1$ and $N_2$ delimiting the swallowtail interior then map to maximum masses for dilute and dense stars, $M_{\rm max}^{\rm dil}=m_a N_1$ and $M_{\rm max}^{\rm den}=m_a N_2$, as functions of $(m_a,f_a)$.
For a given axion model the catastrophe diagram predicts whether galaxies can host only dilute stars, a mixed population of dilute and dense stars, or no long--lived stars at all.

Formation channels such as gravitational collapse of miniclusters and solitons naturally populate the dilute branch.
\cite{Eggemeier2020,Vaquero2019,Visinelli2020,Enander2017,Fairbairn2018,Eggemeier2024,Ellis2022,Eggemeier2023,Schiappacasse2018,Chavanis2025,Painter20242454} Adiabatic accretion drives these objects along the large - $R$ sheet toward the first fold; crossing it triggers collapse and a potential transition to the dense branch, accompanied by significant energy release.
Dense stars may in turn lose mass through axion emission or tidal stripping, moving down the small $R$ sheet toward the second fold, where they can jump back to the dilute branch or disperse.
These transitions correspond to the folds of the swallowtail and are generically associated with non-adiabatic, catastrophic events.

We must note that distinct observable phenomena are associated with particular regions of the catastrophe diagram. These signatures provide valuable insights into the underlying physical processes and can be categorized as follows:
Firstly, in areas populated by stable, dilute stars, one can observe microlensing events and effects caused by substructure. These phenomena occur when the gravitational field of such stars bends and magnifies the light from distant objects, creating detectable variations in brightness. Microlensing is especially useful for probing the existence and distribution of these faint stars, and their subtle gravitational influence reveals details about the structure of the surrounding space.~\cite{SchunckFuchsMielke2006ScalarFieldLenses}
Secondly, in regions where axion stars are dense and actively accreting baryonic matter such as hydrogen axion stars - thermal emission and distinct radio spectral lines may be observed.\cite{bergstrom2013multi} As these axion stars pull in hydrogen, the resulting interactions produce characteristic thermal radiation and radio signals. Detecting these emissions can help identify the presence of axion stars and provide information about their physical properties and the accretion processes involved.\cite{Kouvaris2024,Maseizik2025,Gao2025}
Lastly, transient signals such as radio bursts, gamma-ray flares, and gravitational wave bursts are expected to emerge from explosive events like the collapse or merging of compact objects, especially near and beyond the folds of the catastrophe diagram. These energetic phenomena are brief but intense, and their detection offers crucial evidence for dynamic interactions, including catastrophic collapses or mergers, within these regions.~\cite{Dietrich2019,Roger2016OpticalBosonStar}
Existing and upcoming facilities, from SKA and LOFAR to LSST and LIGO - Virgo - KAGRA, are sensitive to several of these channels.


\section{Discussion and Outlook}

We have shown that the stability and phase structure of axion stars can be organized in terms of fold, cusp, and swallowtail catastrophes of a variational energy functional.  The Gaussian model is approximate, but it captures the essential competition between quantum pressure, gravity, attractive self-interaction, and repulsive higher-order terms.  The main output of the catastrophe-theory description is a global phase diagram: it identifies where stable and unstable branches are created or annihilated and provides analytic critical conditions in the tractable quartic and sextic models.

The revised analysis also clarifies the limitations of this approach.  The variational concavity criterion is a collective-coordinate stability test, not a substitute for a full radial perturbation spectrum.  Similarly, the small-radius behavior of the variational energy is a diagnostic of the potential approximation and not a physical prediction in a regime where the non-relativistic approximation has broken down.  The full cosine potential and the QCD-corrected potential remove unphysical runaway directions introduced by some finite polynomial truncations, but the final fate of dense configurations requires relativistic Einstein--Klein--Gordon simulations.

Several extensions are natural.  The catastrophe-theory analysis can be embedded directly in numerical equilibrium calculations to refine the positions of folds and cusps in physical units.  Rotation, electromagnetic couplings, and multi-field axion sectors may be included as additional control parameters.  Combining the resulting phase diagram with simulations of axion structure formation will allow quantitative predictions for the abundance of dilute and dense axion stars and for the rates of collapse, bosenova-like emission, and compact-object merger signatures.

\section*{Acknowledgement}

This work received institutional funding from Khalifa University. The~authors acknowledge support from the Khalifa University Research and Innovation Grant KU-INT-RIG-2024-046/8474000759, KU-INT-RIG-2023-028, KU-INT-RIG-2024-26, and KU-INT-RIG-2024-053. FVK extends his gratitude to the Team Leader Award under the President’s International Fellowship Initiative of the Chinese Academy of Sciences.

\subsection*{Authors contribution}

All authors equally contributed to the work, All authors have read and agreed to the published version of
the manuscript.
The data analyzed in this study is available at the
GitHub repository.

\bibliographystyle{apsrev4-2}
\bibliography{axionstar}

\end{document}